\documentclass[aps,prl,twocolumn,showpacs,superscriptaddress,groupedaddress,graphics]{revtex4}
\usepackage{graphicx}
\usepackage{dcolumn}
\usepackage{bm}
\usepackage{amssymb}
\usepackage{mwe}
\usepackage{comment}
\usepackage{xcolor}

\begin{document}

\title{Superconducting cavity in a high magnetic field}
\author{Danho Ahn} \affiliation{Center for Axion and Precision Physics Research, Institute for Basic Science, \\Daejeon 34051, Republic of Korea} \affiliation{Department of Physics, Korea Advanced Institute of Science and Technology (KAIST), \\Daejeon 34141, Republic of Korea}
\author{Ohjoon Kwon} \affiliation{Center for Axion and Precision Physics Research, Institute for Basic Science, \\Daejeon 34051, Republic of Korea}
\author{Woohyun Chung} \email[Corresponding author.\\]{gnuhcw@ibs.re.kr} \affiliation{Center for Axion and Precision Physics Research, Institute for Basic Science, \\Daejeon 34051, Republic of Korea}
\author{Wonjun Jang} \affiliation{Center for Quantum Nanoscience, Institute for Basic Science, \\Seoul 33760, Republic of Korea}
\author{Doyu Lee} \altaffiliation{Present address: Samsung Electronics, Hwasung, 18448, Republic of Korea.} \affiliation{Center for Axion and Precision Physics Research, Institute for Basic Science, \\Daejeon 34051, Republic of Korea}
\author{Jhinhwan Lee} \affiliation{Center for Artificial Low Dimensional Electronic Systems, Institute for Basic Science, \\Pohang 37673, Republic of Korea}
\author{Sung Woo Youn} \affiliation{Center for Axion and Precision Physics Research, Institute for Basic Science, \\Daejeon 34051, Republic of Korea}
\author{Dojun Youm} \affiliation{Department of Physics, Korea Advanced Institute of Science and Technology (KAIST), \\Daejeon 34141, Republic of Korea}
\author{Yannis K. Semertzidis} \affiliation{Center for Axion and Precision Physics Research, Institute for Basic Science, \\Daejeon 34051, Republic of Korea} \affiliation{Department of Physics, Korea Advanced Institute of Science and Technology (KAIST), \\Daejeon 34141, Republic of Korea}
\date{\today}

\begin{abstract}
A high Q-factor microwave resonator in a high magnetic field could be used in a wide range of applications, especially for enhancing the scanning speed in axion dark matter research. In this letter, we introduce a polygon-shaped resonant cavity with commercial YBCO tapes covering the entire inner wall. We demonstrated that the maximum Q-factor (TM$_{010}$, 6.93 GHz) of the superconducting YBCO cavity was about 6 times higher than that of a copper cavity and showed no significant degradation up to 8 T at 4 K. This is the first indication of the possible applications of HTS technology to the research areas requiring low loss in a strong magnetic field at high radio frequencies.
\end{abstract}
\pacs{}
\maketitle

\begin{figure}[b]
    \centering
    \includegraphics[width=0.3\textwidth]{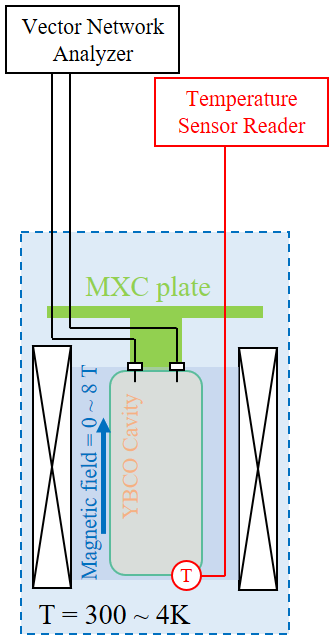}
    \caption{The schematic of the experimental setup. The components inside the blue dashed line are in 4 K which is controlled by a pulse tube. The YBCO cavity is placed inside the bore of an 8 Tesla superconducting magnet which is represented by two white boxes right next to the cavity. The dark blue area shows that the cavity is at the magnetic field center. The range of field strength is zero to 8 Tesla. The red circled "T" represents temperature sensor installed at the bottom of the cavity. The sensor is connected to the temperature sensor reader. The two tiny white rectangles with line segments on the top of the cavity are the antennae which are weakly coupled to the TM$_{010}$ mode. The Q factor is measured by the vector network analyzer with two coaxial cables which are connected to the two antennae.}
    \label{fig:02}
\end{figure}

\begin{figure}[t]
    \centering
    \includegraphics[width=0.45\textwidth]{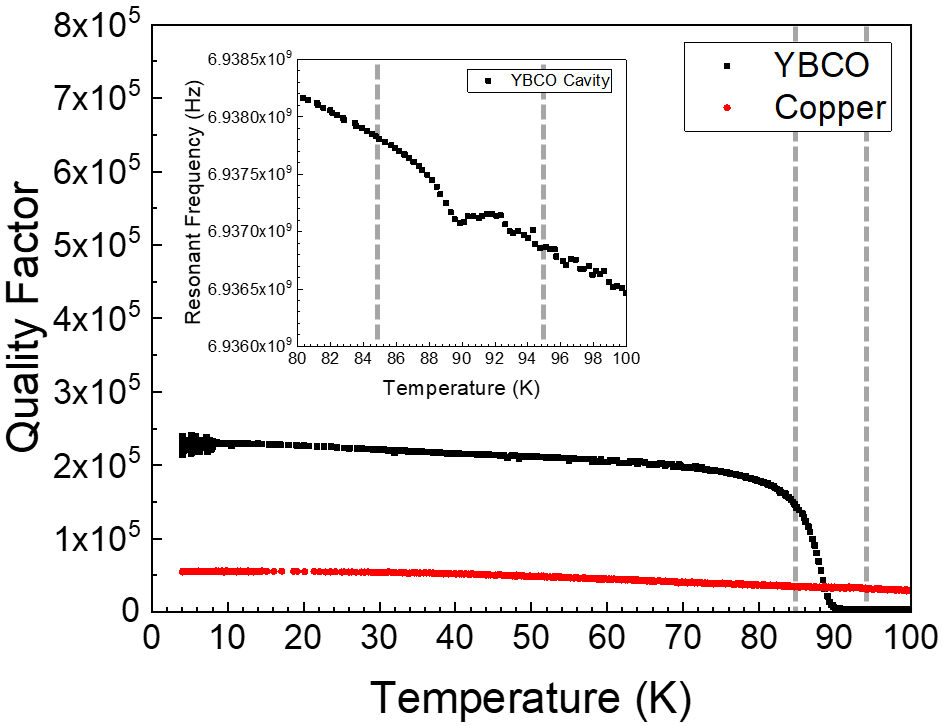}
    \caption{The measurement results of the 12-piece polygon cavities: The Q factor vs. temperature from 4.2 K to 100 K. The black dots are for the YBCO cavity and the red dots are for the copper cavity with the same polygon geometry. The inset plot is the resonant frequency vs. temperature from 80 K to 100 K. The phase transition from normal metal to superconductor starts near 90 K, at which an anomalous frequency shift occurs. The vertical grey dashed lines show the temperatures 85 K and 95 K.}
    \label{fig:03}
\end{figure}

\begin{figure}[t]
    \centering
    \includegraphics[width=0.45\textwidth]{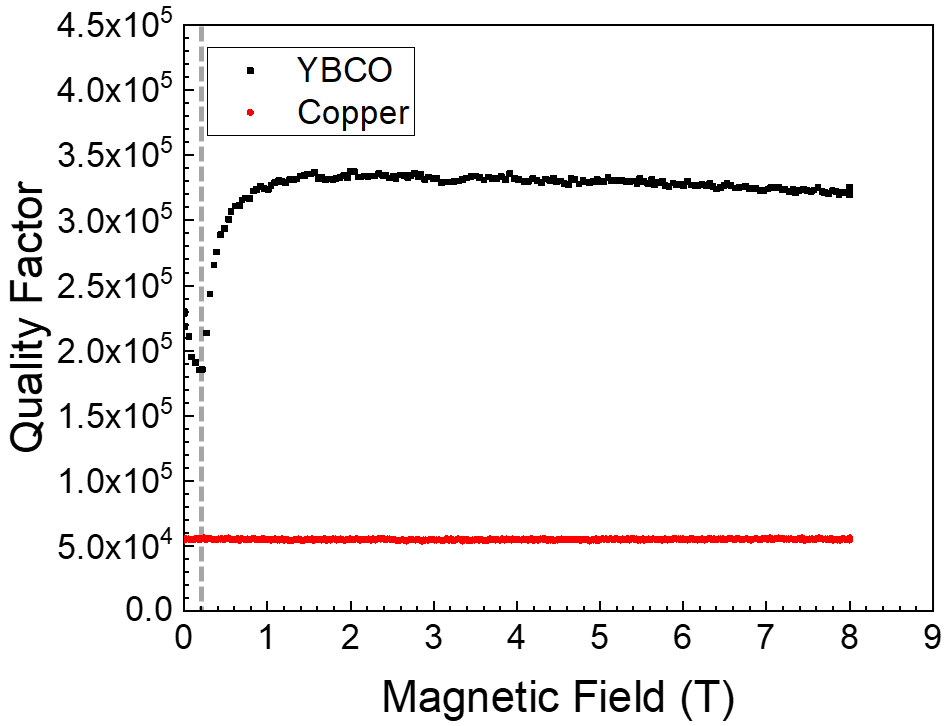}
    \caption{The measurement results of the 12-piece polygon cavities: The Q factor vs. external magnetic field from 0 T to 8 T. The vertical dashed line shows the magnetic field 0.23 T at which the abrupt Q factor enhancement starts. The maximum Q factor is around 330,000. The inset plot is the magnified plot from 0 T to 1 T. }
    \label{fig:04}
\end{figure}

\begin{figure}[t]
    \centering
    \includegraphics[width=0.45\textwidth]{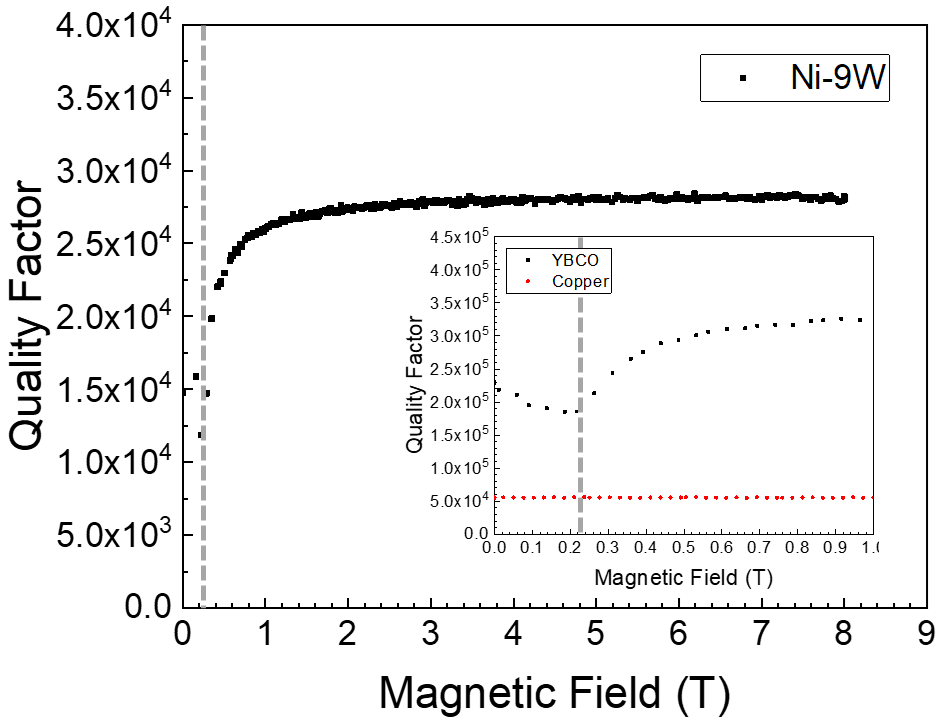}
    \caption{The measurement related to the nickel-tungsten alloy: The Q factor vs. external magnetic field for the 12-piece copper polygon cavity with one piece of Ni-9W (from 0 T to 8 T at 4 K). The Q factor behavior is the same as in the YBCO cavity. There is a abrupt change of the Q factor at 0.23 T.}
    \label{fig:05}
\end{figure}

\begin{figure}[t]
    \centering
    \includegraphics[width=0.45\textwidth]{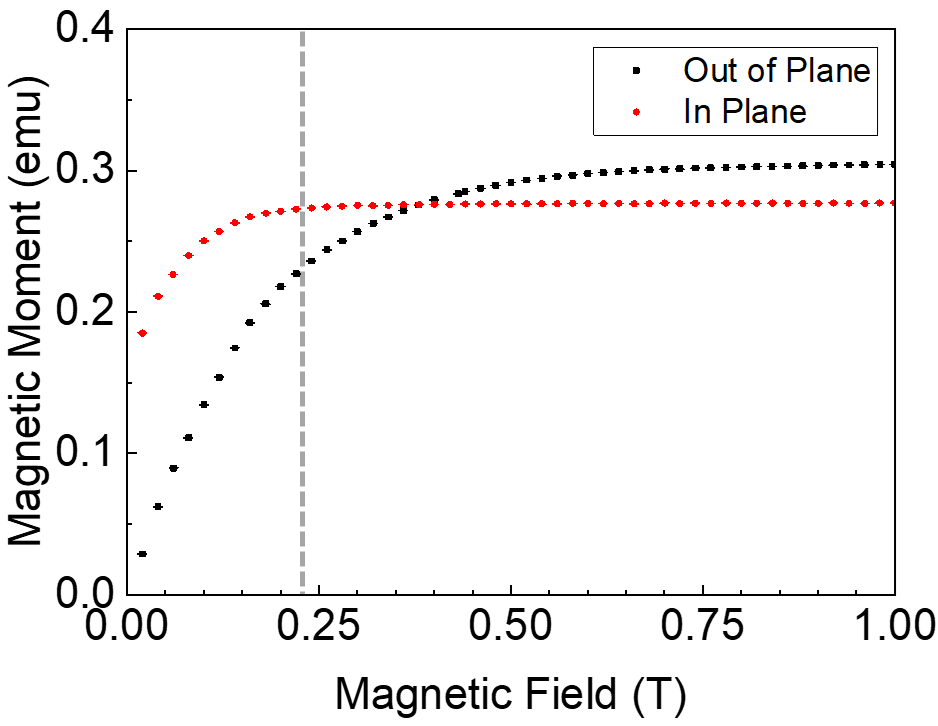}
    \caption{The measurement related to the nickel-tungsten alloy: The magnetization curve for the Ni-9W tape (4mm x 4mm). The magnetic saturation of the Ni-9W tape which is aligned in-plane direction of the DC magnetic field almost ends at 0.23 T at which the Q factor of YBCO cavity is abruptly changed. The vertical dashed line represents 0.23 T. For the case of out-of-plane, the magnetic saturation ends around 1 T at which the Q factor of the YBCO cavity is saturated.}
    \label{fig:06}
\end{figure}

Superconducting radio-frequency (SRF) science and technology involves the application of superconducting properties to radio frequency systems. Due to the ultra-low electrical resistivity, which allows an RF resonator to obtain an extremely high quality (Q) factor, SRF resonant cavities can be used in a broad scope of applications such as particle accelerators \cite{SCcavappli_Accel_01, SCcavappli_Accel_02}, material characterization \cite{SCcavappli_MatChar_01}, and quantum devices \cite{SCcavappli_Qubit_02}. However, the presence of an external magnetic field will destroy the superconducting state above the critical field, which limits scientific productivity in many areas such as high energy particle accelerators \cite{SCcavappli_Accel_03, SCcavappli_Accel_04}, and axion dark matter research \cite{AxSearch_SCcavity_01, AxSearch_CavityExp_01, AxSearch_CavityExp_02, AxSearch_CavityExp_03}. In particular, the axion dark matter detection scheme utilizes a resonant cavity immersed in a strong magnetic field, by which the axions are converted into microwave photons \cite{AxSearch_Sikivie01, AxSearch_Sikivie02}. Maintaining a superconducting cavity in a strong magnetic field will profoundly impact the way axion dark matter experiments are performed. It will substantially increase the searching speed for axions \cite{AxSearch_ScanRate} with expected quality factors of about 10$^{6}$ \cite{AxSearch_AxionQfactor} and will permit the study about the detailed axion signal structure in the frequency domain. Furthermore, achieving  a quality factor more than $10^{6}$ can open a new window for ultra-narrow axion linewidth research \cite{AxSearch_NonVirialized}.\\
\indent The natural choice of material for fabricating the superconducting cavity under a high magnetic field is the high temperature superconductor (HTS) YBa$_2$Cu$_3$O$_{7-x}$ (YBCO) whose surface resistance is lower than copper in any direction of the applied magnetic field. The upper critical field is very high ($>$ 100 T) and the vortex depinning frequency is more than 10 GHz \cite{Copper_Rs_01, YBCO_Propert_Vortex_Golosovsky_01, YBCO_Propert_Vortex_Golosovsky_02}. However, fabricating a 3-D resonant cavity structure with YBCO poses large technical challenges because of YBCO's biaxial texture. Previous studies show that the surface resistance is strongly dependent on the alignment angle between the directions of the YBCO crystal's grain \cite{YBCO_Propert_GrainBoundary} and the applied magnetic field \cite{YBCO_Propert_Vortex_Golosovsky_01}. Moreover, directly forming a grain-aligned YBCO film on the deeply concaved inner surface of the cavities is prohibitively difficult because of the limitations in making the well textured buffer layers and substrate \cite{YBCO_Fabrication_01, YBCO_Fabrication_02, YBCO_Fabrication_03, YBCO_Fabrication_04}.\\ 
\indent A possible solution to this problem is to implement a three-dimensional (3-D) surface with two dimensional (2-D) planar objects. We took advantage of high-grade, commercially available YBCO tapes by AMSC, whose fabrication process, structure, and properties are well-known \cite{YBCO_Tape_01, YBCO_Tape_02}. We chose to use pure YBCO over other rare-earth barium copper oxide (ReBCO) materials which have a high concentration of gadolinium atoms. The RF surface resistance of those ReBCO materials ($\sim$ 1m$\Omega$) \cite{SCcavappli_MatChar_01} could be higher than that of YBCO ($\sim$ 0.1 m$\Omega$) at zero field \cite{YBCO_Propert_Vortex_Golosovsky_02} because gadolinium is paramagnetic, introducing an additional RF energy loss mechanism due to the rotating spins. The substrate and buffer layers of the tape were designed to act as template layers to provide the biaxial texture for the YBCO film. The film architecture of the tape consists of several parts. On the biaxially textured 9 percent nickel-tungsten (Ni-9W) alloy, the 800 nm thick YBCO was deposited on top of the buffer layers which consist of Y$_2$O$_3$, YSZ, and CeO$_2$, and are each 75 nm thick.\\ 
\indent To fabricate a 3-D superconducting cavity utilizing YBCO tapes, we devised a novel scheme by employing a 12-piece polygon cavity to which grain-aligned tapes are attached. Each tape was prepared and attached securely to the inner surface of a cavity piece with a minimum be-
\onecolumngrid
\begin{center}
\begin{figure*}[t]
    \centering
    \includegraphics[width=1\textwidth]{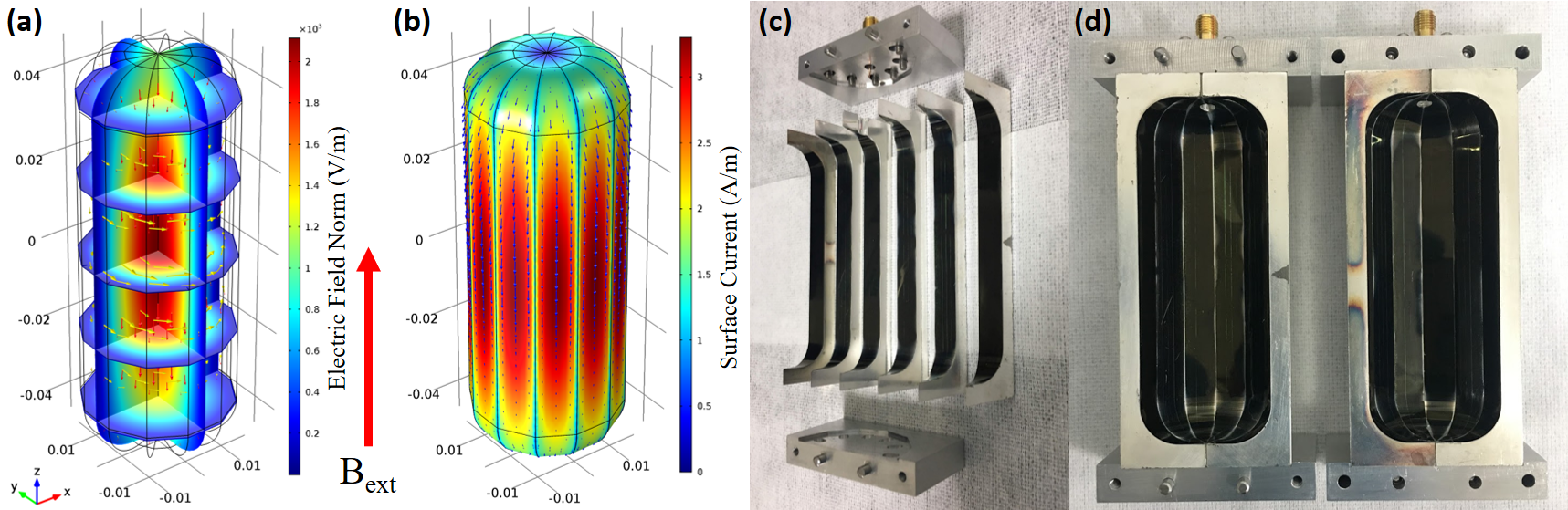}
    \caption{Polygon shape cavity design. (a) The red arrows are electric field lines of the polygon cavity TM$_{010}$ mode from the eigenmode simulation (COMSOL). The yellow arrows are magnetic field lines of the same cavity TM$_{010}$ mode. The color map inside of the cavity represents the amplitude of the electric field of the TM$_{010}$ mode. B$_{ext}$ is the direction of the applied DC magnetic field in the axion cavity experiment. (b) The color map on the surface represents the inner surface current distribution due to the TM$_{010}$ mode. The current flows in the direction of the blue arrows. The surface loss is concentrated in the middle of the cavity. (c) Six aluminum cavity pieces to each of which a YBCO tape is attached. (d) Twelve pieces (two cylinder halves) are assembled to make a whole cavity.}
    \label{fig:01}
\end{figure*}
\end{center}
\twocolumngrid
\noindent nding to prevent cracks (Fig. \ref{fig:01} (a)). An arc radius of 10 mm was applied between the top/bottom and the sidewall surfaces to avoid excess bending stress on the tapes \cite{YBCO_Tape_Bending_01}. The joint mechanism of the twelve separated cavity pieces are designed for accurate alignment of the YBCO tapes upon assembly.  For the fundamental TM modes, most commonly used in axion research, the vertical cuts of the cylindrical cavity do not cause any significant degradation of the Q-factor, since the direction of the surface current in TM$_{010}$ mode and the boundary of each cavity piece are parallel as seen in (Fig. \ref{fig:01} (b)). The results were confirmed by the simulation \cite{Simulation_COMSOL} and the Q-factor measurement of an assembled cavity. The simulation result of the  Q-factor (TM$_{010}$ mode) of the polygon copper cavity with nominal room temperature copper conductivity is 20,400 (Fig. \ref{fig:01} (a), (b)), and the measured Q factor was 19,200. Once the YBCO tape was completely attached to the inner surface of each polygon piece, we removed the protective layers to expose the bare YBCO surface by a novel technique developed at CAPP which is described in the supplementary material. The cut edges of the YBCO tapes exposed on the side were coated by sputtering silver to reduce the RF loss due to small imperfection created in the cutting process. The technique used in this work was optimized for TM modes of a cylindrical cavity but could be applied to any resonators, for minimizing surface losses and resolving contact problems.\\
\indent The assembled cavity was installed in a cryogen-free dilution refrigerator BF-LD400 \cite{Equipment_BF}, equipped with an 8 T cryogen-free NbTi superconductor solenoid \cite{Equipment_Magnet}, and brought to a low temperature of around 4 K. The Q-factor and resonant frequency were measured using a network analyzer through a transmission signal between a pair of RF antennae, which are weakly coupled to the cavity. The coupling strengths of the antennae were monitored throughout the experiment and accounted for in obtaining the unloaded quality factor (Fig. \ref{fig:02}). Measuring the Q factor (TM$_{010}$ mode) of the polygon cavity with the twelve YBCO pieces by varying the temperature, we observed the superconducting phase transition at around 90 K which is in agreement with the critical temperature (T$_c$) of YBCO (Fig. \ref{fig:03}). The global increase of the resonant frequency was due to thermal shrinkage of the aluminum cavity, but an anomalous frequency shift was also observed near the critical temperature. The decrease of the frequency shift at T$_c$ can be attributed to the divergence of the penetration depth of YBCO surface \cite{YBCO_Propert_Vortex_Golosovsky_01}. The maximum Q factor at 4.2 K was about 220,000. The Q-factor for the polygon cavity made of pure (oxygen-free high thermal conductivity, OFHC) copper with the same geometry was measured to be 55,500. Varying the applied DC magnetic field from 0 T to 8 T, at the initial ramping up of the magnet, the Q-factor of the cavity dropped rapidly to 180,000 until the magnetic field reached 0.23 T and then rose up to the maximum value of 335,000, which is about 6 times higher than that of a copper cavity, at around 1.5 T for the TM$_{010}$ mode. From the measurement, we observed that the Q-factor of the resonant cavity's TM$_{010}$ mode did not decrease significantly (changing only a few percent) up to 8 T (Fig. \ref{fig:04}).\\
\indent Investigating the abrupt behavior of the measured Q-factor near 0.23 T, the same Q-factor measurement was repeated using a 12-piece Cu cavity with Ni-9W tape attached only on one piece. The comparison between Fig. \ref{fig:04} and Fig. \ref{fig:05} clearly shows that the unexpected change in Q-factor near 0.23 T is caused by the Ni-9W layer behind the YBCO layer. Furthermore, we also measured the magnetization of a small Ni-9W piece (4 mm x 4 mm) in a magnetic property measurement system Quantum Design MPMS3-Evercool \cite{Equipment_MPMS} with in-plane and out-plane alignments (parallel and perpendicular to the applied magnetic field, respectively) to investigate where the saturation occurs. We used nitric acid to etch the YBCO film on the 4 mm x 4 mm tape. The sample was installed in the equipment with the straw in which the rectangular sample can be aligned in any direction by hand. The measurements show that the magnetic saturation of in-plane (out-of-plane) Ni-9W ends near 0.23 (1.0) T (Fig. \ref{fig:06}). The magnetic saturation lowers the surface resistance, because the atomic spins become more rigid due to the reduction of the magnetic domain walls.
In other words, the Q factor is suddenly increased at 0.23 T because the main surface loss is originated from the side wall which is aligned in the in-plane direction with the external field. After that, other surfaces are saturated until 1.0 T where the Q factor of the YBCO cavity is maximum.\\
\indent The maximum Q-factor achievable with a YBCO cavity is currently unknown but the comparison between the surface resistance of copper at 4 K (5 m$\Omega$ at 5.712 GHz) \cite{Copper_Rs_01} and YBCO at 4 K (0.2 m$\Omega$ at 5-6 GHz) \cite{YBCO_Propert_Vortex_Golosovsky_01} suggests that the Q-factor could be 25 times higher even with a strong magnetic field present. In the near future, improvements are expected for techniques of exposing the bare YBCO surface from the tape, eventually reducing the area where the surface loss occurs inside the cavity. Moreover, if the layer which gives large energy loss, such as Ni-9W, can be eliminated or covered completely, we can expect much higher Q factor. Our design of the vertically split, polygon cavity for implementing biaxially textured YBCO to the inner surface allows us to test the possibility of constructing superconducting resonant cavities which could be used in a strong magnetic field. We demonstrate that it is possible to fabricate a cavity with a YBCO inner surface to maintain a high Q-factor up to 8 T. This result could not only eliminate a significant limitation of SRF applications with a magnetic field in many areas but also provide us with a tool to search for axions even if they are only 10$\%$ of the local dark matter halo.\\

\indent The authors are grateful for the technical advice of Sergey Uchaikin (Magnetic property of YBCO), Junu Jeong (Data Aquisition) at the Center for Axion and Precision Physics Research in the Institute for Basic Science, and Byoungkook Kim at the KAIST Analysis Center for Research Advancement (Magnetic Property Measurement System). This work was supported by IBS-R017-D1-2020-a00 / IBS-IBS-R017-Y1-2020-a00.

\end{document}